\documentstyle[12pt,epsf]{article}
\newcounter{fixy}
\begin{document}
\newenvironment{fixy}[1]{\setcounter{figure}{#1}}
{\addtocounter{fixy}{1}}
\renewcommand{\thefixy}{\arabic{fixy}}
\renewcommand{\thefigure}{\thefixy\alph{figure}}
\setcounter{fixy}{1}

\title{Dilatonic monopoles and ''hairy'' black holes}

\author{{\large Y. Brihaye}$^{\star}$,
{\large B. Hartmann}$^{\diamond}$
and {\large J. Kunz}$^{\diamond}$ \\ \\
$^{\star}${\small Physique-Math\'ematique, Universit\'e de Mons-
Hainaut, Mons, Belgium}\\ \\
$^{\diamond}${\small Fachbereich Physik, Universit\"at Oldenburg,
Postfach 2503, D-26111 Oldenburg, Germany}}

\date{\today}
\newcommand{\dd}{\mbox{d}}\newcommand{\tr}{\mbox{tr}}
\newcommand{\ii}{\mbox{i}}\newcommand{\e}{\mbox{e}}
\newcommand{\pa}{\partial}\newcommand{\Om}{\Omega}
\newcommand{\vep}{\varepsilon}
\newcommand{\bfph}{{\bf \phi}}
\newcommand{\lm}{\lambda}
\def\theequation{\arabic{equation}}
\renewcommand{\thefootnote}{\fnsymbol{footnote}}
\newcommand{\re}[1]{(\ref{#1})}
\newcommand{\bfR}{{\sf R\hspace*{-0.9ex}\rule{0.15ex}%
{1.5ex}\hspace*{0.9ex}}}
\newcommand{\N}{{\sf N\hspace*{-1.0ex}\rule{0.15ex}%
{1.3ex}\hspace*{1.0ex}}}
\newcommand{\Q}{{\sf Q\hspace*{-1.1ex}\rule{0.15ex}%
{1.5ex}\hspace*{1.1ex}}}
\newcommand{\C}{{\sf C\hspace*{-0.9ex}\rule{0.15ex}%
{1.3ex}\hspace*{0.9ex}}}
\newcommand{\eins}{1\hspace{-0.56ex}{\rm I}}
\renewcommand{\thefootnote}{\arabic{footnote}}

\maketitle
\begin{abstract}
We study gravitating monopoles and non-abelian black holes
of SU(2) Einstein-Yang-Mills-Higgs
theory coupled to a massless dilaton.
The domain of existence of these solutions is bounded and
decreases with increasing dilaton coupling strength.
The critical solutions of this system are Einstein-Maxwell-dilaton
solutions.
\end{abstract}
\medskip
\medskip

\newpage
\section{Introduction}
In SU(2) Einstein-Yang-Mills-Higgs (EYMH) theory, with the Higgs field
in the adjoint representation,
globally regular gravitating magnetic monopole solutions
\cite{ewein,bfm,lw}
emerge smoothly from the flat space 't Hooft-Polyakov monopole \cite{thooft},
when gravity is coupled.
Since the mass of the non-abelian monopole in flat space is proportional to the
Higgs vaccum expectation value $v$,
while the size of the monopole core is inversely proportional to $v$,
the monopole size should become comparable to the Schwarzschild radius,
when $v$ becomes sufficiently large, and the non-abelian monopole
should implode to a black hole \cite{frie}.
Thus regular magnetic monopoles should exist only in a limited
domain of the EYMH parameter space.

Indeed, as revealed by numerical analysis \cite{ewein,bfm,lw},
regular gravitating monopoles exist only up to a maximal value
$\alpha_{\rm max}$ of the coupling constant $\alpha$
(which is proportional the Higgs vacuum expectation value $v$
and the inverse Planck mass).
Beyond $\alpha_{\rm max}$ the only solutions with unit magnetic charge
are embedded abelian solutions,
namely Reissner-Nordstr\o m (RN) black hole solutions.
Thus $\alpha_{\rm max}$ limits the domain of the EYMH parameter
space, where non-abelian monopole solutions exist.

Besides the maximal value of $\alpha$, $\alpha_{\rm max}$,
the critical value of $\alpha$, $\alpha_{\rm cr}$,
is of particular importance.
As $\alpha$ approaches the critical value
$\alpha_{\rm cr} \le \alpha_{\rm max}$ \cite{foot0},
the non-abelian monopole solutions approach
a critical solution with a degenerate horizon.
Outside the horizon the critical solution corresponds
to an extremal RN solution with unit charge,
while it is non-singular inside the horizon.
Thus at $\alpha_{\rm cr}$,
the non-abelian monopole branch bifurcates with the branch
of extremal RN black holes \cite{ewein,bfm,lw}.

Distinct from embedded RN black holes,
genuine non-abelian magnetically charged black hole solutions
emerge from the globally regular magnetic monopole solutions,
when a finite regular event horizon is imposed
\cite{ewein,bfm,aichel}.
Representing counterexamples to the ``no-hair'' conjecture for black holes,
these black hole solutions may be viewed as
``black holes within monopoles'' \cite{ewein},
and have been interpreted as
bound monopole-black hole systems \cite{ash}.

The domain of existence in parameter space
of genuine non-abelian black hole solutions is also bounded \cite{ewein,bfm}.
In particular, for a fixed value of the coupling constant $\alpha$
($\alpha < \alpha_{\rm max}$),
the domain of existence of non-abelian black hole solutions
is bounded by a maximal value of the horizon radius
$x_{h,\rm max}$ (which depends on $\alpha$),
and only relatively small non-abelian black hole solutions exist,
as predicted on general grounds \cite{sud}.

Here we investigate the influence of a massless dilaton
on the gravitating non-abelian monopole solutions,
since dilatons appear naturally in many unified theories,
including string theory.
The influence of a massless dilaton
on the non-abelian monopole solutions in flat space
has remarkable similarities
to the influence of gravity on the monopole solutions \cite{ymhd}.
In particular, in flat space the non-abelian dilatonic monopole solutions
exist only below a maximal value of the coupling constant
$\gamma$
(which is proportional the Higgs vacuum expectation value $v$
and the dilaton coupling constant).
As $\gamma$ approaches a critical value,
the non-abelian dilatonic monopole solutions approach
a critical solution, which corresponds to
an embedded abelian dilatonic monopole solution.
Thus the branch of regular non-abelian monopole solututions
bifurcates with the branch of singular abelian monopole solutions \cite{ymhd}.

Considering the influence of both gravity and a dilaton, we expect
that the domain of existence in parameter space of
the non-abelian monopoles in Einstein-Yang-Mills-Higgs-dilaton (EYMHD) theory
should also be bounded.
This expectation is in distinction to arguments from string theory,
which suggest, that regular non-abelian dilatonic monopole solutions exist
for any value of the coupling constant \cite{hl}.
We furthermore expect that
the non-abelian dilatonic monopole solutions should
also show a critical behaviour, when the respective
coupling constants are varied.
In particular, the branch of gravitating
non-abelian dilatonic monopoles should bifurcate with the corresponding
branch of embedded abelian solutions,
namely Einstein-Maxwell-dilaton (EMD) solutions \cite{emd}.

We also investigate the influence of a dilaton
on the non-abelian black hole solutions,
and we determine their domain of existence in parameter space.
Furthermore, we obtain simple relations between
the dilaton field and the metric, and the dilaton charge and the mass
for the monopoles and black holes \cite{kks-long}.

We review the EYMHD action in section II, where we discuss
the spherically symmetric ansatz, the field equations,
the boundary conditions and the relations
between metric and dilaton field.
We discuss our numerical results for the globally regular
monopole solutions in section III and for the
non-abelian black hole solutions in section IV.
The conclusions are presented in section V.
We briefly review the numerical procedure in Appendix A.

\section{SU(2) Einstein-Yang-Mills-Higgs-Dilaton Equations of Motion}

\subsection{SU(2) Einstein-Yang-Mills-Higgs-dilaton action}
We consider the action of the EYMHD theory
\begin{equation}
S=S_{G}+S_{M}=\int L_{G}\sqrt{-g}d^{4}x+ \int L_{M}\sqrt{-g}d^{4}x
\ . \end{equation}
The gravity Lagrangian $L_{G}$ is given by
\begin{equation}
4\pi L_{G}=\frac{1}{4G}R
\ , \end{equation}
where $G$ is Newton`s constant.
The matter Lagrangian is given in terms of the gauge fields ${A_{\mu}}^a$,
the Higgs fields ${\Phi}^a$, and the dilaton field $\Psi$ ($a=1,2,3$)
\begin{equation}
4\pi L_{M}=-\frac{1}{4} e^{2\kappa\Psi}F_{\mu\nu}^{a}F^{\mu\nu,a}
-\frac{1}{2}\partial_{\mu}\Psi\partial^{\mu}\Psi
-\frac{1}{2}D_{\mu}\Phi^{a} D^{\mu}\Phi^{a}-e^{-2\kappa\Psi}V(\Phi^{a})
\ , \end{equation}
with Higgs potential
\begin{equation}
V(\Phi^{a})=\frac{\lambda}{4}(\Phi^{a}\Phi^{a}-v^2)^2
\ , \end{equation}
The non-abelian field strength tensor is given by
\begin{equation}
F_{\mu\nu}^{a}=\partial_{\mu}A_{\nu}^{a}-\partial_{\nu}A_{\mu}^{a}+
g\varepsilon_{abc}A_{\mu}^{b}A_{\nu}^{c}
\ , \end{equation}
and the covariant derivative of the Higgs field in the adjoint representation
by
\begin{equation}
D_{\mu}\Phi^{a}=\partial_{\mu}\Phi^{a}+
g\varepsilon_{abc}A_{\mu}^{b}\Phi^{c}
\ , \end{equation}
$g$ denotes the gauge field coupling constant,
$\kappa$ the dilaton coupling constant,
$\lambda$ the Higgs field coupling constant,
and $v$ the vacuum expectation value of the Higgs field.

\subsection{Spherically symmetric ansatz}

To construct static spherically symmetric dilatonic monopoles and black holes,
we employ for the metric the static spherically symmetric Ansatz
with Schwarzschild-like coordinates
\begin{equation}
ds^{2}=-A^{2}(r)N(r)dt^2+N^{-1}(r)dr^2+r^2 d\theta^2+r^2\sin^2\theta
d^2\varphi
\ , \end{equation}
with the metric functions $A(r)$ and
\begin{equation}
N(r)=1-\frac{2m(r)}{r}
\ , \end{equation}
where the metric function $m(r)$ is the mass function.
For the gauge and Higgs fields, we use the purely magnetic hedgehog ansatz
\cite{thooft,ewein,bfm}
\begin{equation}
{A_r}^a={A_t}^a=0
\ , \end{equation}
\begin{equation}
{A_{\theta}}^a=- \frac{1-K(r)}{g} {e_{\varphi}}^a
\ , \ \ \ \
{A_{\varphi}}^a= \frac{1-K(r)}{g}\sin\theta {e_{\theta}}^a
\ , \end{equation}
\begin{equation}
{\Phi}^a=v H(r) {e_r}^a
\ . \end{equation}
A non-vanishing time-component of the gauge field would lead to dyon
solutions \cite{jul,bhk}.

The dilaton is a scalar field depending only on $r$
\begin{equation}
\Psi=\Psi(r)
\ . \end{equation}

\subsection{Field equations}

Let us now introduce dimensionless coordinates and fields
\begin{equation}
x=rgv \ , \ \ \mu=mgv \ ,\ \ \phi=\frac{\Phi}{v}\ , \ \ \psi=\frac{\Psi}{v}
\label{scale}
\ . \end{equation}
The Lagrangian and the resulting set of differential equations
then depend only on three
dimensionless coup\-lings constants, $\alpha$, $\beta$ and $\gamma$,
\begin{equation}
\alpha =\sqrt{G}v =\frac{M_W}{gM_{\rm Pl}} \ , \ \
\beta=
 \frac{\sqrt{\lambda}}{g} = \frac{M_H}{\sqrt{2}M_W} \ , \ \
\gamma =\kappa v =\frac{\kappa M_W}{g}
\ , \end{equation}
where $M_W=g v$, $M_H= \sqrt{2\lambda} v$ and $M_{\rm Pl}=1/\sqrt{G}$.
Special cases are $\kappa=\sqrt{G}$ (i.e.~$\gamma=\alpha$),
corresponding to string theory,
and
$\kappa^2= G\frac{(2+n)}{n}$
(i.e.~$\gamma=\sqrt{\frac{2+n}{n}}\alpha$),
corresponding to $(4+n)$-dimensional Kaluza-Klein theory \cite{emd}.

Inserting the Ansatz into the Lagrangian and varying with respect
to the matter fields yields the Euler-Lagrange equations
\begin{equation}
(e^{2\gamma\psi}ANK')'=A(e^{2\gamma\psi}\frac{K(K^2-1)}{x^2}+H^2 K)
\ , \label{dgl1} \end{equation}
\begin{equation}
(x^2 ANH')'=AH(2K^2+ \beta^2 x^2 e^{-2\gamma\psi}(H^2-1))
\ , \label{dgl2} \end{equation}
\begin{equation}
(x^2 AN\psi')'=2\gamma A(e^{2\gamma\psi}(N(K')^2+\frac{(K^2-1)^2}{2 x^2})-
e^{-2\gamma\psi} \frac{\beta^2 x^2}{4}(H^2-1)^2)
\ , \label{dgl3} \end{equation}
where the prime denotes the derivative with respect to $x$.\\
Variation with respect to the metric yields the Einstein
equations
\begin{equation}
G_{\mu\nu}=2\alpha^2 T_{\mu\nu}
\ , \end{equation}
where $T_{\mu\nu}$ is the energy-momentum tensor
\begin{equation}
4\pi T_{\mu\nu}=g_{\mu\nu}L_{M}-2\frac{\partial L_{M}}{\partial g^{\mu\nu}}
\ . \end{equation}
With the following combinations of the Einstein equations
\begin{equation}
G_{tt}=2\alpha^2 T_{tt}=-2\alpha^2 A^2 N L_{M}
\ , \end{equation}
\begin{equation}
g^{rr}G_{rr}-g^{tt}G_{tt}=-4\alpha^2 N \frac{\partial L_{M}}{\partial N}
\ , \end{equation}
we obtain for the metric functions $\mu$ and $A$
the differential equations
$$
\mu ' = \alpha^2 \left(e^{2\gamma\psi}N(K')^2 + \frac{1}{2}N x^2(H')^2+
\frac{1}{2x^2}(K^2-1)^{2} e^{2\gamma\psi}+K^2 H^2 \right. \\ \nonumber
$$
\begin{equation}
 +  \left. \frac{\beta^{2}}{4}x^2
(H^2-1)^2 e^{-2\gamma\psi}+\frac{1}{2}Nx^{2}(\psi ')^2 \right)
\ , \label{dgl4} \end{equation}
\begin{equation}
A'=\alpha^2 x A \left(\frac{2(K')^2}{x^2}e^{2\gamma\psi}+(H')^2+(\psi ')^2
\right)
\ . \label{dgl5} \end{equation}
For $\gamma\equiv 0$, the dilaton decouples,
$\psi\equiv 0$,
and the set of equations
reduces to the EYMH equations studied previously \cite{ewein,bfm,lw}.

\subsection{Boundary conditions}

For the system of three second order equations, Eqs.~(\ref{dgl1})-(\ref{dgl3}),
and two first order equations, Eqs.~(\ref{dgl4})-(\ref{dgl5}),
we need to specify eight boundary conditions.

\subsubsection{Dilatonic monopoles}

Let us first consider the boundary conditions for
static globally regular solutions with finite energy,
which are asymptotically flat.
Regularity at the origin requires the boundary conditions
\begin{equation}
K(0)=1 \ , \ \ H(0)=0 \ , \ \ \partial_{x}\psi|_{x=0}=0 \ , \ \ \mu(0)=0
\ . \label{bc1} \end{equation}
At infinity asymptotic flatness and finite energy impose
the boundary conditions
\begin{equation}
K(\infty)=0 \ , \ \ H(\infty)=1 \ , \ \ \psi(\infty)=0 \ , \ \ A(\infty)=1
\ , \label{bc2} \end{equation}
where the condition on $\psi$ can be imposed because of dilatational symmetry
\cite{foot1}, and the condition on $A$ fixes the time coordinate.
Note, that the value of the metric function $\mu$ at infinity
determines the dimensionless mass of the solutions,
$\mu(\infty)/\alpha^2$.

\subsubsection{Dilatonic black holes}

To construct black hole solutions
which possess a regular event horizon at radius $x_h$,
corresponding boundary conditions must be imposed at the horizon.
Regularity at the horizon requires for the matter functions
\begin{equation}
N'K'=\frac{K(K^2-1)}{x^2}+e^{-2\gamma\psi}H^2 K
\ , \label{bc4} \end{equation}
\begin{equation}
N'H'=\frac{H}{x^2}(2K^2+ \beta^2 x^2 e^{-2\gamma\psi}(H^2-1))
\ , \label{bc5} \end{equation}
\begin{equation}
N'\psi'=\frac{2\gamma}{x^2}(e^{2\gamma\psi}(N(K')^2+\frac{(K^2-1)^2}{2 x^2})-
e^{-2\gamma\psi}\frac{\beta^2 x^2}{4}(H^2-1)^2)
\ , \label{bc6} \end{equation}
and for the metric functions
\begin{equation}
\mu(x_h)=\frac{x_h}{2}
\ , \label{bc3} \end{equation}
and $A(x_h)< \infty$.
At infinity, the black hole solutions satisfy the same boundary
conditions as the globally regular solutions,
Eqs.~(\ref{bc2}).

\subsection{Relations between metric functions and dilaton field}

In Einstein-Yang-Mills-dilaton (EYMD) theory, relations
between the metric functions and dilaton field are known \cite{kks-long}.
In the following we demonstrate, that corresponding relations
hold in EYMHD theory.

Let us introduce the abbreviations
\begin{equation}
E_A = N(K')^2+{(K^2-1)^2\over {2x^2}} \ , \ \
E_V = \beta^2{x^2\over 4} (H^2-1)^2  \ ,
\nonumber \end{equation}
\begin{equation}
E_H = {1\over 2} N x^2(H')^2+K^2H^2 \ , \ \
F = e^{2\gamma\psi}
\ . \end{equation}
In terms of these the dilaton equation and the first Einstein equation
read,
\begin{equation}
(x^2AN\psi')' = 2\gamma A (FE_A-F^{-1}E_V)
\ , \end{equation}
\begin{equation}
\mu' = \alpha^2 (FE_A + E_H + F^{-1} E_V + {1\over 2} N x^2(\psi')^2)
\ . \end{equation}
Contraction of the Einstein equations yields for the
curvature scalar $R$
\begin{equation}
-R =2\alpha^2\left( -N(\psi')^2-{4\over{x^2}} F^{-1} E_V-{2\over{x^2}} E_H
\right)
\ . \end{equation}
Combination of these equations gives \cite{kks-long}
\begin{equation}
(x^2 AN \psi')' = \frac{2\gamma A}{\alpha^2} ({\mu'} -{1\over 4} x^2R)
\ . \end{equation}
With
\begin{equation}
 A x^2 R = - \left[ x^2 (2 A' N + A N') \right]' + 4 A \mu'
\ , \label{R1} \end{equation}
we obtain for the dilaton field the equation
\begin{equation}
(AN x^2 \psi')' =  \frac{\gamma }{2 \alpha^2}
  \left( x^2 (2 A' N + A N') \right)'
\ . \label{rel1} \end{equation}
This equation can be integrated, yielding
\begin{equation}
 \psi' =  \frac{\gamma }{2 \alpha^2}
  \left( \ln ( A^2 N) \right)' + \frac{C}{A N x^2}
\ , \label{rel2} \end{equation}
with integration constant $C$.

We can now derive the global relations
between mass $\mu(\infty)/\alpha^2$
and dilaton charge $D$, defined via
\begin{equation}
\psi(x) \stackrel {x\rightarrow \infty} {\longrightarrow}
-\frac{D}{x}
\ . \label{bc8} \end{equation}
Considering first the regular solutions,
we integrate Eq.~(\ref{rel1}) from zero to infinity,
and take into account the asymptotic behaviour
of the dilaton field, Eq.~(\ref{bc8}),
and of the metric functions
\begin{equation}
A'(x)\propto O(x^{-3}), \ \
N'(x)\propto \frac{2\mu(\infty)}{x^2}+O(x^{-3})
\ , \end{equation}
as well as $N(\infty)=A(\infty)=1$.
This yields the relation between the dilaton charge and the mass of
the solution \cite{kks-long}
\begin{equation}
D = \gamma\frac{\mu(\infty)}{\alpha^2}
\ . \label{res1} \end{equation}
Consequently the integration constant $C$ in Eq.~(\ref{rel2})
vanishes for regular solutions.
Integrating Eq.~(\ref{rel2})
from $x$ to infinity then gives a simple relation
between dilaton field and $tt$-component of the metric
\cite{kks-long}
\begin{equation}
\psi(x) =  \frac{\gamma}{ 2 \alpha^2} \ln( A^2 N)
   = \frac{\gamma}{2 \alpha^2} \ln(-g_{tt})
\ . \label{res2} \end{equation}

For black hole solutions
we integrate Eq.~(\ref{rel1}) from the horizon to infinity.
With the appropriate boundary conditions we thus obtain for black holes
the relation \cite{kks-long}
\begin{equation}
D = \frac{\gamma}{\alpha^2}
  \left(\mu(\infty)-\frac{1}{2}x^2 A N' |_{x_{\rm H}}
  \right)
\ . \label{res3} \end{equation}
The integration constant $C$ therefore does not vanish for black holes.

\section{Dilatonic Monopoles}

The non-abelian magnetic monopole solutions of
Yang-Mills-Higgs (YMH) theory \cite{thooft} persist
in EYMH theory \cite{ewein,bfm,lw}, when gravity is coupled,
and in YMHD theory \cite{ymhd},
when a dilaton field is coupled,
for small values of the respective coupling constants.
In both cases, the solutions bifurcate with an abelian solution
at a critical value of the respective coupling constant.

Here we construct non-abelian magnetic monopoles, coupled to both gravity
and a dilaton field. We solve the set of differential equations
Eqs.~(\ref{dgl1})-(\ref{dgl3}) and Eqs.~(\ref{dgl4})-(\ref{dgl5})
numerically (see Appendix A for details),
subject to the boundary conditions Eqs.~(\ref{bc1})-(\ref{bc2}).
We show, that the gravitating dilatonic non-abelian magnetic monopole solutions
also bifurcate with an abelian solution, an extremal EMD solution \cite{emd}.

\subsection{Monopoles in EYMH theory}

Let us begin by briefly recalling the EYMH results for gravitating monopoles
obtained in \cite{ewein,bfm,lw},
corresponding to the limit $\gamma=0$ of EYMHD theory.

For a fixed value of the Higgs self-coupling $\beta$,
a branch of gravitating non-abelian monopole solutions
emerges smoothly from the flat space monopole solution,
when $\alpha$ is increased.
This branch exists on the interval $0 \le \alpha \le \alpha_{\rm max}$.
We refer to this branch as main branch.
The maximal value of $\alpha$, $\alpha_{\rm max}$,
decreases with increasing $\beta$.

For vanishing and small values of $\beta$,
a short second branch of non-abelian monopole solutions
is present on the interval
$\alpha_{\rm cr} \le \alpha \le \alpha_{\rm max}$,
and bifurcates with the main branch at $\alpha_{\rm max}$.
We refer to this branch as secondary branch.
At $\alpha_{\rm cr}$
the secondary branch bifurcates with a branch of embedded abelian solutions,
the extremal RN solutions of Einstein-Maxwell (EM) theory,
carrying unit magnetic charge and mass
$\mu_{\rm RN}(\infty)/\alpha^2=1/\alpha$.
For larger values of $\beta$, the secondary branch disappears,
and $\alpha_{\rm cr}=\alpha_{\rm max}$ \cite{ewein,bfm}.
Increasing $\beta$ further leads to a new phenomenon \cite{lw,foot2}.

Let us now consider, how the solutions approach the critical solution
as $\alpha \rightarrow \alpha_{\rm cr}$, while $\beta$ is kept fixed.
The metric function $N(x)$ possesses a minimum, which decreases monotonically
as $\alpha \rightarrow \alpha_{\rm cr}$.
At $\alpha_{\rm cr}$
the minimum of $N(x)$ reaches zero at the value of the radial coordinate
$x=x_h= \alpha_{\rm cr}$.
For $x \ge x_h$, the metric function $N(x)$ of the critical solution
is identical to the metric function
$N_{\rm RN}(x)= (1-\alpha_{\rm cr}/x)^2$ of the extremal RN solution,
whereas it is different from the abelian function for $x \le x_h$.
$x_h$ represents the event horizon of the RN black hole.

Likewise, the matter field functions $K(x)$ and $H(x)$
approach the constant values $K_{\rm RN}(x)=0$ and $H_{\rm
RN}(x)=1$
of the extremal RN solution
as $\alpha \rightarrow \alpha_{\rm cr}$ for $x \ge x_h$,
whereas they tend to non-trivial functions for $x \le x_h$.
Thus the critical solution reached at $\alpha_{\rm cr}$
is identical to an extremal RN solution outside the
radius $x_h$, while it retains distinct non-abelian features
inside the radius $x_h$.

\subsection{Monopoles in YMHD theory}

Let us now recall the YMHD results for
dilatonic non-abelian monopoles in flat space \cite{ymhd},
corresponding to the limit $\alpha=0$ of EYMHD theory.
The dilatonic non-abelian monopoles in YMHD theory have many features
in common with the non-abelian monopoles in curved space.

When the dilaton is coupled,
the main branch of dilatonic non-abelian monopoles
emerges smoothly from the 't Hooft-Polyakov monopole,
when $\gamma$ is increased, while
the Higgs self-coupling $\beta$ is kept fixed.
The main branch extends up to a maximal value of $\gamma$, $\gamma_{\rm max}$.
For vanishing and small values of $\beta$, a secondary branch exists
on the interval $\gamma_{\rm cr} \le \gamma \le \gamma_{\rm max}$,
and bifurcates with the main branch at $\gamma_{\rm max}$.
At $\gamma_{\rm cr}$
the secondary branch bifurcates with a branch of embedded abelian
monopole solutions.
For larger values of $\beta$, the secondary branch disappears,
and $\gamma_{\rm cr}=\gamma_{\rm max}$ \cite{ymhd}.

Intriguingly, even the numerical value for the maximal
coupling constant $\gamma_{\rm max}$ is almost identical
to the maximal coupling constant $\alpha_{\rm max}$ (for $\beta=0$).

\subsection{Monopoles in EYMHD theory}

Let us now consider how the presence of both gravity and a massless
dilaton field affects the non-abelian monopoles.
Generalizing the observations from EYMH and YMHD theory,
we expect, that non-abelian monopoles in EYMHD theory
also show a critical behaviour, when the respective
coupling constants $\alpha$ and/or $\gamma$ are varied.
In particular, the emerging branch of gravitating
non-abelian dilatonic monopoles should bifurcate with the corresponding
branch of embedded abelian solutions.
Here these should be Einstein-Maxwell-dilaton (EMD) solutions \cite{emd}.
Thus the domain of existence in parameter space of
the gravitating dilatonic non-abelian monopoles should be bounded.

\boldmath
\subsubsection{$\beta=0$}
\unboldmath

We first discuss the non-abelian monopoles of EYMHD theory for $\beta=0$.

Starting from the flat space dilatonic monopole solution
for some value of the coupling constant $\gamma$,
$\gamma \le \gamma_{\rm cr}$ \cite{ymhd},
we couple gravity by choosing a small but finite value of
the coupling constant $\alpha$.
Increasing $\alpha$, we obtain the main branch of
dilatonic non-abelian monopole solutions,
which exists on the interval $0 \le \alpha \le \alpha_{\rm max}$.
The short secondary branch of dilatonic non-abelian monopole solutions,
present on the interval $\alpha_{\rm cr} \le \alpha \le \alpha_{\rm max}$,
bifurcates with the main branch at $\alpha_{\rm max}$.
At $\alpha_{\rm cr}$
the secondary branch bifurcates with a branch of embedded abelian solutions,
which correspond to extremal EMD solutions,
carrying unit magnetic charge and mass
\begin{equation}
\frac{\mu_{\rm EMD}(\infty)}{\alpha^2}=
 \frac{1}{\sqrt{\alpha^2+\gamma^2}}
\ . \label{massemd} \end{equation}
This is in complete accordance with our expectation \cite{foot3}.

Let us now consider in more detail how the non-abelian EYMHD monopole
solutions approach the critical value of $\alpha$,
when $\gamma$ is held fixed (and $\beta=0$).
As in the EYMH system, the metric function $N(x)$ of the EYMHD system
develops a minimum that decreases
monotonically along the main and secondary branch.
However, along the secondary branch the area close to the minimum
starts to form a plateau,
which broadens as $\alpha \rightarrow \alpha_{\rm cr}$.
At the same time
the function $N(x)$ becomes increasingly steep close to the origin.
In the limit $\alpha \rightarrow \alpha_{\rm cr}$,
the plateau extends up to $x=0$,
and the non-abelian function $N(x)$ converges to the
function $N_{\rm EMD}(x)$ of the extremal EMD solution
with unit magnetic charge \cite{emd},
and with the same values of $\alpha$ and $\gamma$.
But since the value of the extremal EMD solution at the origin
is given by \cite{emd},
\begin{equation}
N_{\rm EMD}(0)= \left( \frac{\gamma^2}{\alpha^2 + \gamma^2} \right)
\ , \end{equation}
while the boundary conditions for the non-abelian solution require
$N(0)=1$, convergence is pointwise.
In contrast to the EYMH case, the minimum of the function $N(x)$
thus does not reach zero in the limit $\alpha \rightarrow \alpha_{\rm cr}$,
when a dilaton field is present, so no horizon forms.

We illustrate the dependence of the metric function $N(x)$
of the EYMHD monopoles on the coupling strength $\alpha$
for $\gamma=0.5$ in Fig.~1a.
The metric function $N(x)$ is shown for
the maximal value of $\alpha$, $\alpha_{\rm max}=1.30702$,
and for $\alpha = 1.29459$, a value very close to the critical
value $\alpha_{\rm cr}$, where a pronounced plateau is already visible.
For comparison, the extremal EMD solution for $\alpha = 1.29459$
is also shown.

The matter functions $K(x)$ and $H(x)$ also converge to their
constant EMD values, $K_{\rm EMD}(x)=0$ and $H_{\rm EMD}(x)=1$,
in the limit $\alpha \rightarrow \alpha_{\rm cr}$,
Because of their respective boundary conditions at the origin,
$K(0)=1$ and $H(0)=0$, convergence to the EMD solution is again pointwise.

Likewise, the dilaton function $\psi(x)$ approaches
the EMD dilaton function
\begin{equation}
\psi_{\rm EMD}(x)=\frac{\gamma}{\alpha^2+\gamma^2}\ln(1-\frac{X_-}{X})
\ , \label{psiemd} \end{equation}
with
\begin{equation}
x/\alpha=X(1-\frac{X_{-}}{X})^{\gamma^2/(\alpha^2+\gamma^2)} \ , \ \
X_-= \left( \frac{\alpha^2 + \gamma^2}{\alpha^2} \right)^{\frac{1}{4}}
\ . \label{psiemd2}\end{equation}
Since $X=X_-$ corresponds to the origin $x=0$,
$\psi_{\rm EMD}(x)$ tends logarithmically
to minus infinity for $x\rightarrow 0$.
We illustrate the dilaton function $\psi(x)$ for $\gamma=0.5$ in Fig.~1b
for $\alpha_{\rm max}=1.30702$ and $\alpha = 1.29459$,
i.~e.~for the same parameter values as in Fig.~1a.
Convergence of the non-abelian functions $\psi(x)$
to the abelian function $\psi_{\rm EMD}(x)$, also shown in Fig.~1b,
is clearly seen.
In particular,
the value of the dilaton function at the origin, $\psi(0)$,
diverges for $\alpha \rightarrow \alpha_{\rm cr}$,
in accordance with Eqs.~(\ref{psiemd})-(\ref{psiemd2}).

For a fixed value of $\gamma$,
the mass $\mu(\infty)/\alpha^2$ of the non-abelian solutions
decreases with increasing $\alpha$.
In Fig.~2 we exhibit the dependence of the mass
of the non-abelian monopoles on the coupling constant $\alpha$.
In particular, we exhibit the mass of the EYMH solutions for $\gamma=0$ \cite{bfm}
and the mass of the EYMHD solutions for $\gamma=1.0$ and $1.2$.
Also shown is the mass of the respective abelian solutions,
namely the extremal RN solutions for $\gamma=0$
and the extremal EMD solutions for finite $\gamma$.
The figure nicely illustrates the bifurcation
of the non-abelian solutions with the corresponding abelian solutions.
(The short secondary branches are very close to the abelian branches.)

Fig.~2 also shows, that for a fixed value of $\alpha$,
the mass $\mu(\infty)/\alpha^2$ of the non-abelian solutions
decreases with increasing $\gamma$.
Interestingly, the mass of the critical solution, where the
bifurcation takes place appears to be rather constant,
as seen in Fig.~2.
Given by Eq.~(\ref{massemd}), the mass of the critical solution
depends only on the combination $\alpha^2+\gamma^2$.
Since $\alpha_{\rm max}$ depends on $\gamma$, we conclude, that
\begin{equation}
\left( \alpha_{\rm max}(\gamma) \right)^2 + \gamma^2  \approx \rm const
\ . \end{equation}
As demonstrated in Fig.~3,
the values $\alpha_{\rm max}(\gamma)$ lie indeed approximately on
a circle, $\sqrt{\alpha_{\rm max}^2 + \gamma^2 } \approx 1.4$,
which forms the boundary of the domain of existence of the
non-abelian EYMHD monopole solutions.

When varying $\gamma$ while keeping $\alpha$ fixed,
we obtain consistent results.
In particular, for fixed $\alpha$, the main branch
of dilatonic non-abelian monopole solutions
exists on the interval $0 \le \gamma \le \gamma_{\rm max}$,
while the short secondary branch is
present on the interval $\gamma_{\rm cr} \le \gamma \le \gamma_{\rm max}$.
At $\gamma_{\rm cr}$
the secondary branch bifurcates with
the corresponding branch of extremal EMD solutions \cite{foot4}.

Thus the dilaton field affects the gravitating non-abelian monopoles
in a number of ways. Varying $\gamma$ along the main and
secondary branch while keeping $\alpha$ fixed,
the non-abelian core of the monopole solution shrinks
and at the same time, the mass decreases.
At the critical value of $\gamma$ the core shrinks to zero size,
and the mass reaches the value of an extremal EMD solution,
representing a naked singularity.
This critical behaviour differs distinctly from the EYMH case,
where the exterior part of the critical solution
coincides with an abelian extremal black hole,
and the interior part of the critical solution
retains the non-abelian features.

One should keep in mind the scale invariance of EYMHD theory, though.
The above discussion of the critical behaviour corresponds
to the choice $\psi(\infty)=0$
for the value of the dilaton field at infinity, Eq.~(\ref{bc2}).
This implies that the value of the dilaton field at the origin
diverges for $\gamma \rightarrow \gamma_{\rm cr}$,
$\psi(0) \rightarrow -\infty$.
A rescaling of the solutions \cite{foot1}
such that $\psi(0)=0$
should yield a critical solution with a regular interior
and a finite size monopole core,
but with an asymptotically diverging dilaton field and a diverging mass
\cite{ymhd,lm}.

\boldmath
\subsubsection{$\beta \ne 0$}
\unboldmath

We have performed a detailed analysis
of the EYMHD monopole solutions for small values of $\beta$.
Here the features observed for $\beta=0$ persist.
The domain of existence shrinks with increasing $\beta$.
But its boundary remains close to a circle for small values of $\beta$,
becoming slightly deformed with increasing $\beta$.
This is illustrated in Fig.~3 for $\beta=1$ and $\beta=2$.

While we have made an extensive analysis of the EYMHD monopole solutions
for vanishing and small $\beta$, we have not been able to cover much of the
parameter space for larger values of $\beta$.
When trying to determine the domain of existence in the $\alpha$-$\gamma$-plane
for larger values of $\beta$,
we have encountered numerical problems \cite{foot5}.
Here new phenomena appear \cite{foot6},
which make the numerical analysis extremely difficult.
Certainly further investigation of the non-abelian EYMHD monopole
solutions for large values of $\beta$ is called for,
but employment of another numerical procedure might be necessary.

\section{Dilatonic Black Holes}

Non-abelian black holes in EYMHD theory have many features in common with
non-abelian black holes in EYMH theory.
In particular, they are limited in size and they tend to critical
abelian solutions at critical values of the horizon radius.

The black hole solutions of EYMHD theory depend on four parameters,
$\alpha$, $\beta$, $\gamma$ and the horizon radius $x_h$.
Here we limit the discussion to $\beta = 0$.
We construct the black hole solutions,
by solving the set of differential equations
Eqs.~(\ref{dgl1})-(\ref{dgl3}) and
Eqs.~(\ref{dgl4})-(\ref{dgl5})
numerically (see Appendix A for details),
subject to the boundary conditions Eqs.~(\ref{bc2})-(\ref{bc6}),
thus the black hole solutions are asymptotically flat
and they possess a regular horizon.

\subsection{Black holes in EYMH theory}

Let us again briefly recall some properties of the
non-abelian black holes in EYMH theory \cite{ewein,bfm,aichel,lw}.
For a fixed value of the coupling constant $\alpha$
($\alpha < \alpha_{\rm max}$),
black hole solutions emerge from the globally regular solution,
when a finite horizon radius is imposed.
They persist up to a maximal value of the horizon radius
$x_{h,\rm max}$, depending on $\alpha$.
Thus the domain of existence of genuine non-abelian black hole solutions
is bounded, and only relatively small non-abelian black hole solutions exist
\cite{sud}.

We dinstiguish two regions of black holes solutions
with different critical behaviour.
For a fixed value of $\alpha > \sqrt{3}/2$,
the critical solution approached corresponds to an extremal RN solution.
In contrast, for a fixed value of $\alpha < \sqrt{3}/2$,
the critical solution approached corresponds to a non-extremal RN solution.
In particular, for values of $\alpha$ not too close to
$\alpha = \sqrt{3}/2$, two branches of solutions appear,
a main branch $0 \le x_h \le x_{h,\rm max}$
and a secondary branch $x_{h,\rm cr} \le x_h \le x_{h,\rm max}$.

\subsection{Black holes in EYMHD theory}

Let us first consider
the domain of existence of the non-abelian black hole solutions.
We recall, that the domain of existence
of the non-abelian EYMHD monopole solutions
is approximately bounded by a circle
in the $\alpha$-$\gamma$-plane.
Imposing a small but fixed horizon radius $x_h$, the domain of
existence of the non-abelian EYMHD black hole solutions
decreases with respect to the domain of existence of the regular solutions.
With increasing size of the horizon, the domain of existence
of the black hole solutions decreases further.
Since the maximal value of $\gamma$,
$\gamma_{\rm max}(\alpha=0)$ decreases stronger
with increasing $x_h$ than the
maximal value of $\alpha$, $\alpha_{\rm max}(\gamma=0)$,
the domain of existence of the non-abelian black hole solutions
is no longer approximately bounded by a circle
in the $\alpha$-$\gamma$-plane for finite values of $x_h$.
This is demonstrated in Fig.~4, which shows the domain of existence
in the $\alpha$-$\gamma$-plane
for the non-abelian black hole solutions
with horizon radii $x_h=0.1$ and $x_h=0.3$,
in addition to the domain of existence of
the non-abelian monopole solutions.

In the $\alpha$-$x_h$-plane
the domain of existence of the non-abelian black hole solutions
decreases with increasing $\gamma$,
as demonstrated in Fig.~5 for $\gamma=0$, $\gamma=0.5$ and $\gamma=1.0$
\cite{foot7}.
In the EYMH system, the domain of existence
is separated into two distinct regions by
the straight line $\alpha=x_h$,
representing the extremal RN solutions.
In particular, the maximum of the boundary curve
$x_{h,\rm max}(\alpha)$
occurs on this line for $\alpha = \sqrt{3}/2$.
As seen in Fig.~5, the boundary of the domain of existence
in the $\alpha$-$x_h$-plane changes drastically with increasing $\gamma$.
The boundary curve becomes rather smooth and the maximum disappears.
We note, that in contrast to the EYMH case,
in the EYMHD case abelian solutions exist for
arbitrarily small values of the horizon radius $x_h$,
since extremal EMD solutions have $x_h=0$.

While Fig.~5 demonstrates the dependence of the maximal value
of $\alpha$, $\alpha_{\rm max}$, on $x_h$ for fixed $\gamma$,
we now consider the dependence of the critical value
of $\alpha$, $\alpha_{\rm cr}$,
on $x_h$ for fixed $\gamma$.
In the following we restrict the discussion to $\gamma=0.5$.

For fixed $x_h$ and increasing $\alpha$,
the main branch of solutions extends from $\alpha=0$ to
$\alpha_{\rm max}(x_h)$.
At $\alpha_{\rm max}(x_h)$ the main branch bifurcates with the
secondary branch.
Depending on the value of $x_h$,
we observe two qualitatively different types of behaviour
for the secondary branch.
When $0 \leq x_h(\gamma=0.5) \leq 0.22$, the secondary branch
ends at a finite critical value of $\alpha$, whereas when
$0.22 \leq x_h(\gamma=0.5) \leq 0.48$,
the secondary branch extends all the way to $\alpha=0$.
This is demonstrated in Fig.~6, where the values of the matter functions
at the horizon, $K(x_h)$ and $H(x_h)$, are shown for $x_h=0.2$
and $x_h=0.3$.

When decreasing $\alpha$ along the secondary branch
for $0 \leq x_h(\gamma=0.5) \leq 0.22$,
it first appears, that the solution
tends to an extremal EMD solution on the interval $x > x_h$
with a divergent derivative of the metric function $N$ at $x_h$.
Indeed, the mass ratio $\mu(\infty)/\mu_{\rm EMD}(\infty)$,
remains close to one on a large part of the secondary branch,
as seen in Fig.~7.
Decreasing $\alpha$ further, however, indicates
that the critical solution is not an extremal
EMD solution, but a non-extremal EMD solution.
While for $\alpha \rightarrow \alpha_{\rm cr}$
the matter functions $K$ and $H$ approach their abelian values,
$K(x_h)\rightarrow 0$ and $H(x_h)\rightarrow 1$,
as seen in Fig.~6 for $x_h=0.2$,
the mass ratio $\mu(\infty)/\mu_{\rm EMD}(\infty)$,
tends to a value significantly greater than one,
as seen in Fig.~7.
The profiles of the metric function $N$
and the dilaton function $\psi$ support the conclusion as well,
that the critical solution is a non-extremal EMD solution.
In particular, the plateau of the metric function $N$, that forms when
$\alpha$ is decreased from $\alpha_{\rm max}$,
disappears again, when $\alpha$ is decreased further
towards $\alpha_{\rm cr}$.

For $0.22 \leq x_h(\gamma=0.5) \leq 0.48$
no critical value of $\alpha$ exists, instead
the secondary branch extends
all the way to $\alpha=0$, where $K(x_h)$ and $H(x_h)$
reach intermediate values in the interval $]0:1[$
in the limit $\alpha\rightarrow 0$,
as demonstrated in Fig.~6 for $x_h=0.3$.
The corresponding mass ratio $\mu(\infty)/\mu_{\rm EMD}(\infty)$
for $x_h=0.3$ is shown in Fig.~7.

\section{Conclusions}

We have constructed non-abelian monopoles and black holes in EYMHD theory,
for vanishing and small Higgs self-coupling.
Like the non-abelian monopole solutions of EYMH theory,
the non-abelian monopole solutions of EYMHD theory reach critical solutions,
when a critical value of the Higgs vacuum expectation value is reached.
Therefore, the domain of existence of non-abelian monopole solutions
is bounded also in EYMHD theory.

As the critical solution is approached,
the branch of non-abelian EYMH monopole solutions bifurcates
with a branch of embedded abelian solutions,
corresponding to extremal RN solutions with unit charge.
In the presence of a massless dilaton field,
the branch of non-abelian EYMHD monopole solutions also bifurcates
with a branch of embedded abelian solutions,
which here correspond to extremal EMD solutions with unit charge.
Thus without the dilaton field the critical solution is associated with
a black hole solution, whereas in the presence of the dilaton field
the critical solution is associated with a naked singularity.

Like the non-abelian black hole solutions of EYMH theory,
the non-abelian black hole solutions of EYMHD theory
are limited in size. Only relatively small black hole solutions exist
\cite{sud}.
The critical solutions reached
at a critical value of the horizon radius $x_h$
also are associated with embedded abelian solutions,
corresponding to RN black hole solutions in EYMH theory
and to EMD black hole solutions in the presence of a dilaton field.

The boundary of the domain of existence of the non-abelian
monopole solutions is approximately a circle
in the $\alpha$-$\gamma$-plane.
For non-abelian black hole solutions the domain of existence
decreases with increasing horizon radius $x_h$
in the $\alpha$-$\gamma$-plane.
In the $\alpha$-$x_h$-plane the domain of existence decreases
with increasing $\gamma$, exhibiting a much smoother boundary line
for finite values of $\gamma$ than for $\gamma=0$.

Besides presenting numerical results we have also derived an
analytical formula relating the dilaton function and the metric
in the case of the dilatonic monopoles,
and we have given analytical relations between
the mass and the dilaton charge for monopoles and black holes.

We have not yet considered the radial excitations of
the dilatonic monopoles and black holes, which should be present,
in analogy to the radial excitations of the EYMH solutions
\cite{ewein,bfm}.

EYMH theory also possesses static axially symmetric
monopole and black hole solutions, which are not spherically symmetric
\cite{hkk}. These black hole solutions, in particular, show,
that Israels theorem cannot be extended to EYMH theory
\cite{ewein2}.
Currently, we are investigating such static axially symmetric
solutions also in EYMHD theory.

\noindent
{\bf Acknowledgements:} One of us (B.H.) acknowledges the Belgium
F.N.R.S. for financial support.

\section{Appendix A: Numerical Procedure}

To construct the dilatonic monopole and black hole solutions numerically,
we employ a collocation method for boundary-value ordinary
differential equations developed by Ascher, Christiansen and Russell
\cite{COLSYS}.
The set of non-linear coupled differential equations
Eqs.~(\ref{dgl1}) -(\ref{dgl3}) and
Eqs.~(\ref{dgl4}) -(\ref{dgl5})
is solved using the damped
Newton method of quasi-linearization.
At each iteration step a linearized problem
is solved by using a spline collocation at Gaussian points.
Since the Newton method works very well,
when the initial approximate solution is close to the true solution,
the dilatonic monopole and black hole solutions for varying parameters
$\alpha$, $\beta$, $\gamma$ and $x_h$ are obtained by continuation.

The linearized problem is solved on a sequence of meshes
until the required accuracy is reached.
For a particular mesh
$x_i=x_{1} < x_{2} <...< x_{N+1}=x_o$,
where $x_i$ and $x_o$ are the boundaries of the interval,
and $h_i = x_{i+1}-x_i$,
$h= {\rm max}_{1\le i\le N} h_i$,
a collocation solution $\vec{v}$ $(x)=(v_{1},v_{2},...v_{d})$
is determined. Each component $v_{n}(x)\in C^{m_{n}-1}[x_i,x_o]$,
is a polynomial of degree smaller than $k+m_{n}$,
where $m_{n}$ is the order of the $n$-th equation,
and $k$ is an integer bigger than the highest order
of any of the differential equations.
The collocation solution is required
to satisfy the set of differential equations
at the $k$ Gauss-Legendre points in each subinterval
as well as the set of boundary conditions.

When approximating the true solution $u_{n}(x)$
by the collocation solution $v_{n}(x)$,
an error estimate in each subinterval $x \in [x_{i},x_{i+1})$
is obtained from the expression
\begin{equation}
|| u_{n}^{(l)}(x)-v_{n}^{(l)}(x)|| _{(i)}=c_{n,l}
| u_{n}^{(k+m_{n})}(x_{i})| h_{i}^{k+m_{n}-l}+O(h^{k+m_{n}-l+1})
\nonumber \end{equation}
\begin{equation}
l=0, ...., m_{n}-1 \ , \ \ \ n=1,...,d \ ,
\label{error} \end{equation}
where $c_{n,l}$ are known constants.
Also, using Eq.~(\ref{error}) a redistribution of the mesh points is performed
to roughly equidistribute the error.
With this
adaptive mesh selection procedure, the equations are solved on a sequence of
meshes until the successful stopping criterion is reached,
where the deviation of the collocation solution from the
true solution is below a prescribed error tolerance \cite{COLSYS}.

For the numerical solutions, we typically specified
the error tolerance in the range $10^{-4}-10^{-6}$.
The number of mesh points used in these calculations
was typically about $350$.
We calculated the dilatonic monopole solutions
on the finite interval $[0,x_o]$
and the black hole solutions
on the finite interval $[x_h,x_o]$,
where $x_h$ represents the horizon of the black hole,
with $x_o$ ranging from $10^2$ to $10^4$.

\vfill
\newpage

\begin{fixy}{0}
\begin{figure}
\mbox{\epsffile{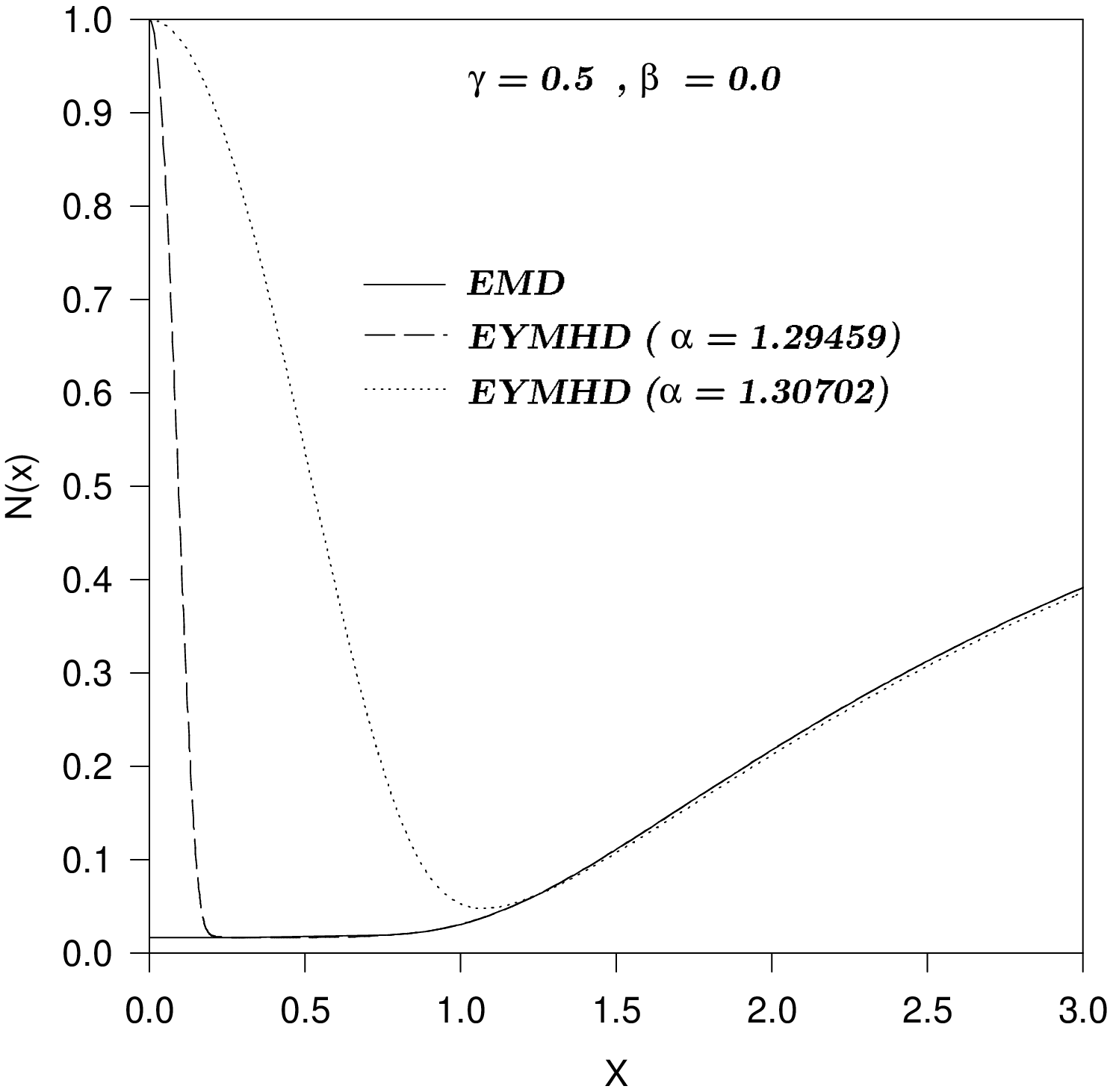}}
\caption{The metric function $N(x)$ of the non-abelian monopole solutions
is shown for $\gamma=0.5$ and two values of $\alpha$,
the maximal value of $\alpha$, $\alpha_{\rm max}=1.30702$,
and $\alpha = 1.29459$, a value very close to the critical
value $\alpha_{\rm cr}$.
For comparison, the metric function $N_{\rm EMD}(x)$
of the extremal EMD solution for $\gamma=0.5$ and $\alpha = 1.29459$
is also shown.}
\end{figure}

\newpage
\begin{figure}
\mbox{\epsffile{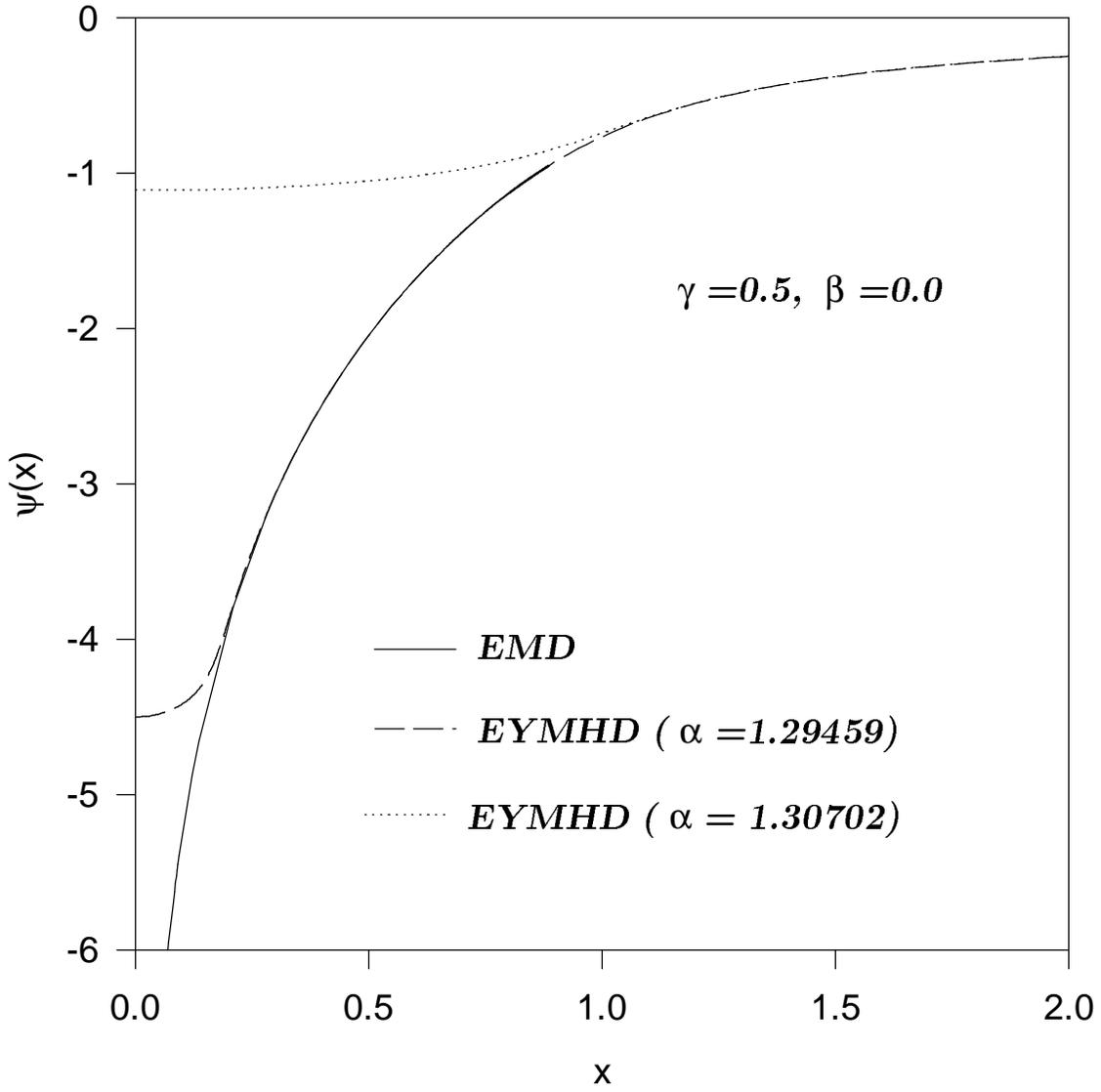}}
\caption{Same as Fig.~1a for the dilaton function $\psi(x)$.}
\end{figure}
\end{fixy}

\newpage
\begin{fixy}{-1}
\begin{figure}
\mbox{\epsffile{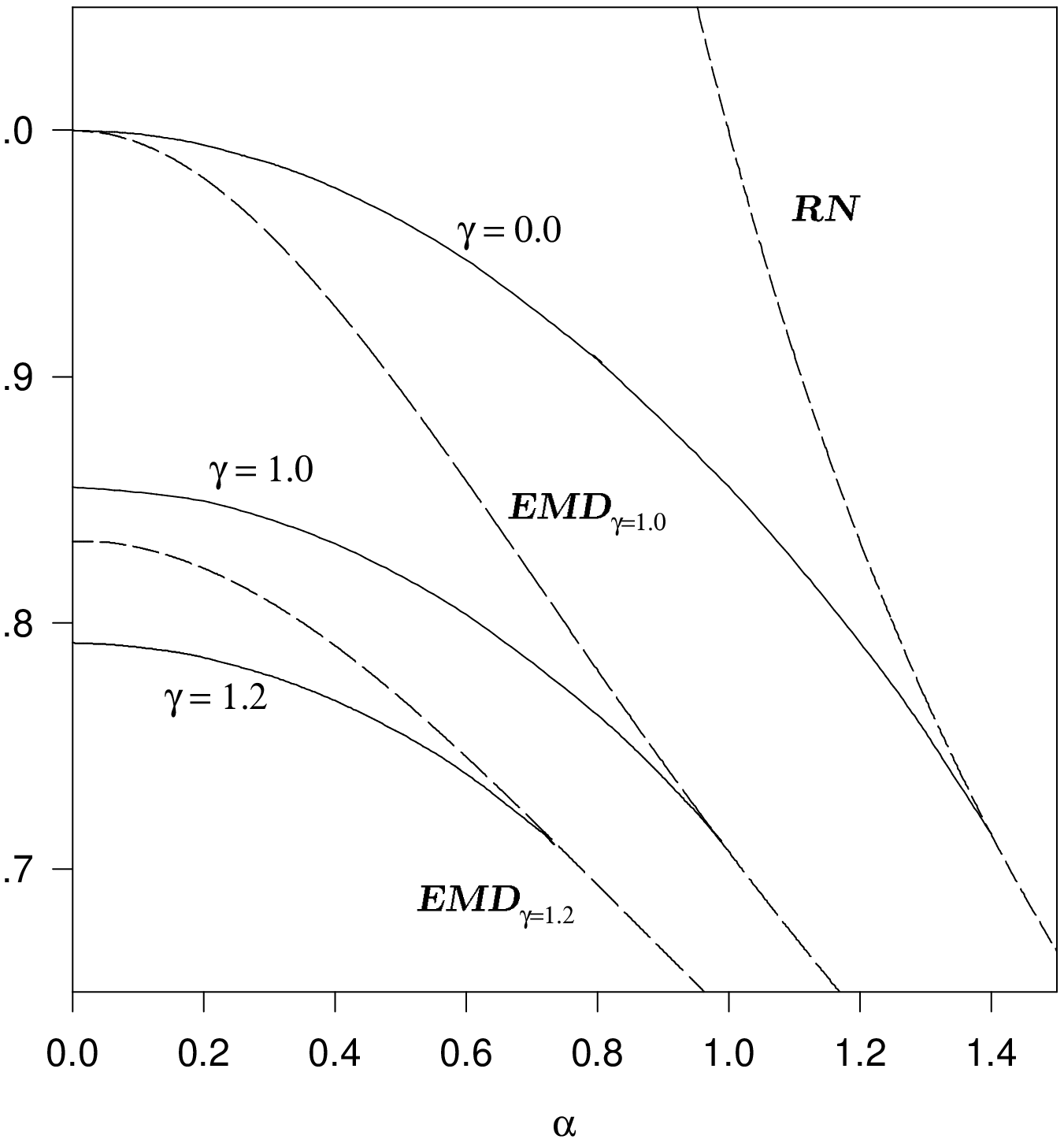}}
\caption{The dependence of the mass $\mu(\infty)/\alpha^2$ on $\alpha$
is shown for the non-abelian monopole solutions
for three different values of $\gamma$,
$\gamma=0$, $\gamma=1$ and $\gamma=1.2$.
For comparison, the dependence of the mass on $\alpha$
is also shown
for the corresponding abelian solutions (RN and EMD, respectively).}\
\end{figure}
\end{fixy}

\newpage
\begin{fixy}{-1}
\begin{figure}
\mbox{\epsffile{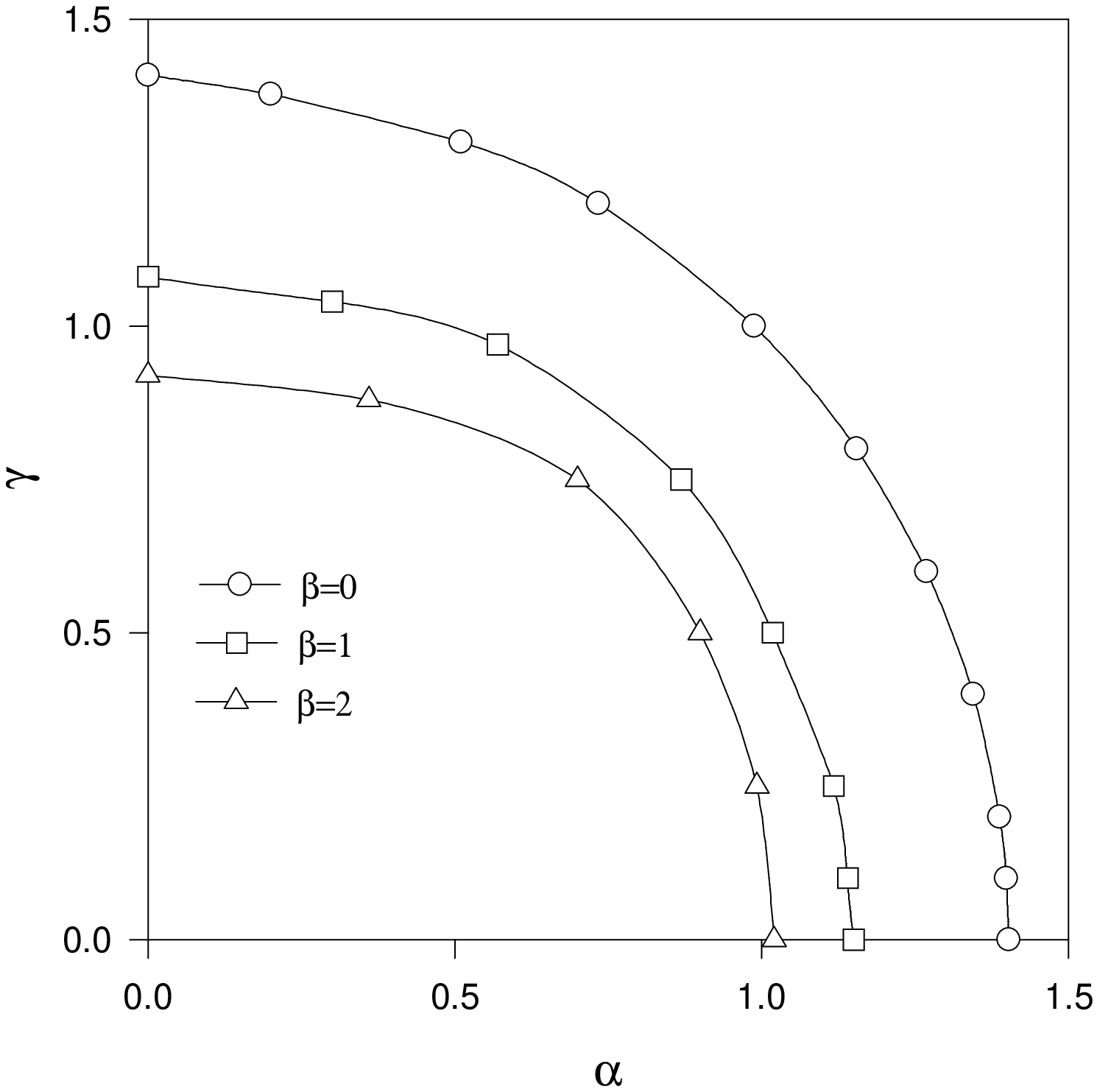}}
\caption{The domain of existence of the non-abelian monopole solutions
in the $\alpha$-$\gamma$-plane is shown for three values
of the Higgs self-coupling,
$\beta=0$, $\beta=1$ and $\beta=2$.}
\end{figure}
\end{fixy}

\newpage
\begin{fixy}{-1}
\begin{figure}
\mbox{\epsffile{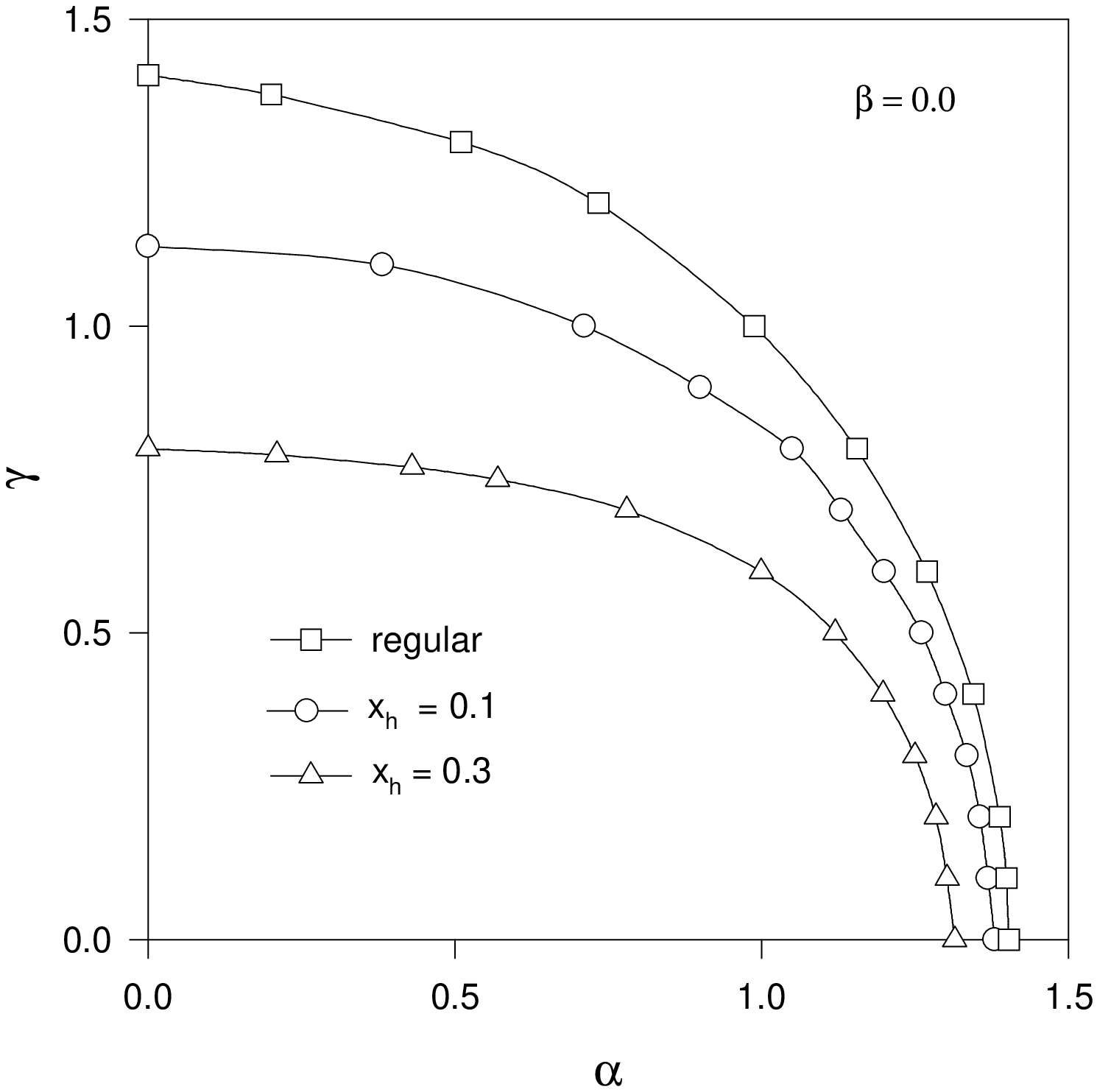}}
\caption{The domain of existence of the non-abelian black hole solutions
in the $\alpha$-$\gamma$-plane is shown for $x_h=0.1$ and $x_h=0.3$.
Also shown is
the domain of existence of the non-abelian monopole solutions.}
\end{figure}
\end{fixy}

\newpage
\begin{fixy}{-1}
\begin{figure}
\mbox{\epsffile{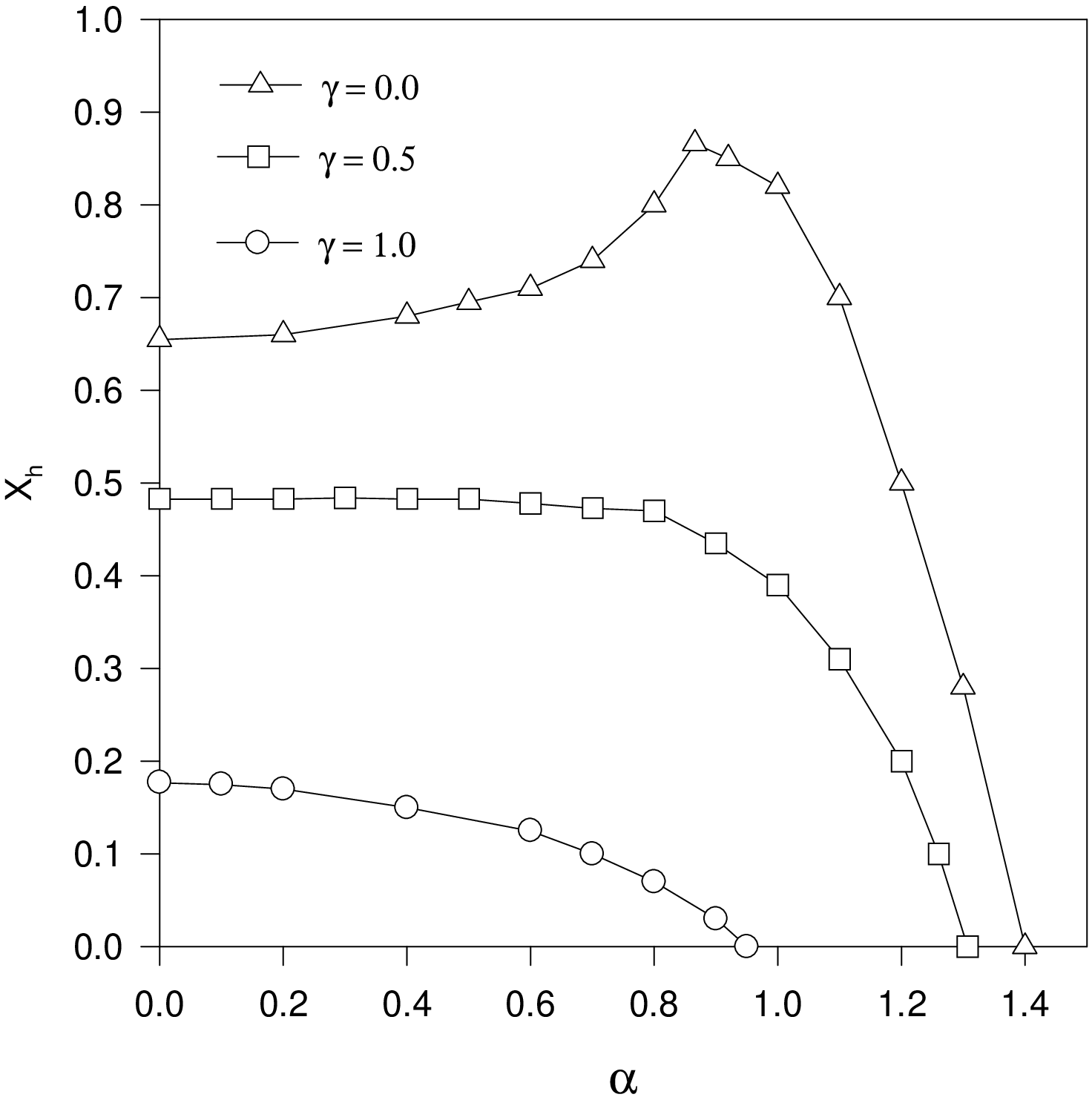}}
\caption{The domain of existence of the non-abelian black hole solutions
in the $\alpha$-$x_h$-plane is shown for three values of the dilaton
coupling constant, $\gamma=0$, $\gamma=0.5$ and $\gamma=1$.}
\end{figure}
\end{fixy}

\newpage
\begin{fixy}{-1}
\begin{figure}
\mbox{\epsffile{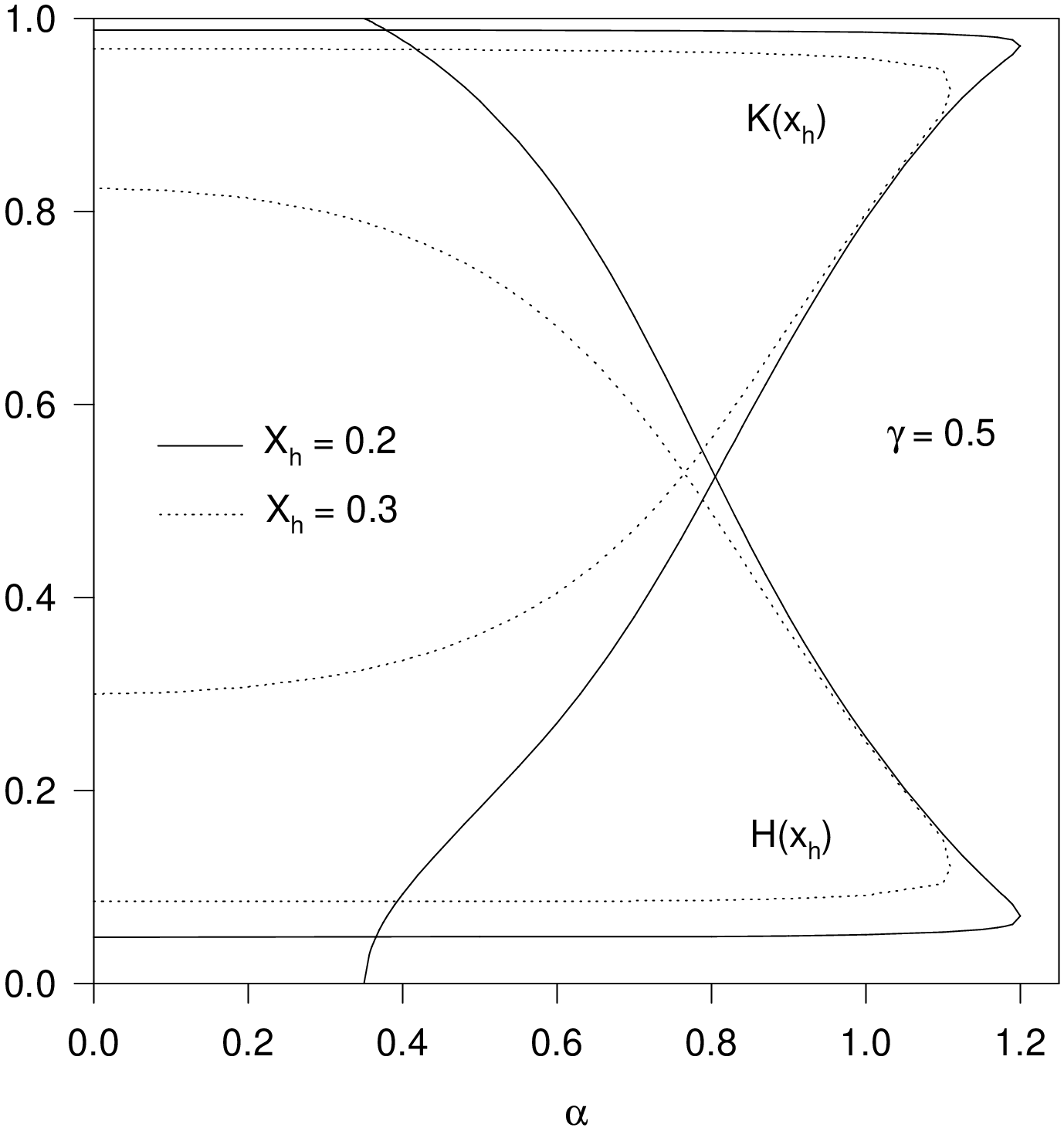}}
\caption{The values of the matter functions at the horizon,
$K(x_h)$ and $H(x_h)$, of the non-abelian
black hole solutions are shown for
$\gamma=0.5$ and for $x_h=0.2$ and $x_h=0.3$.}
\end{figure}
\end{fixy}

\newpage
\begin{fixy}{-1}
\begin{figure}
\mbox{\epsffile{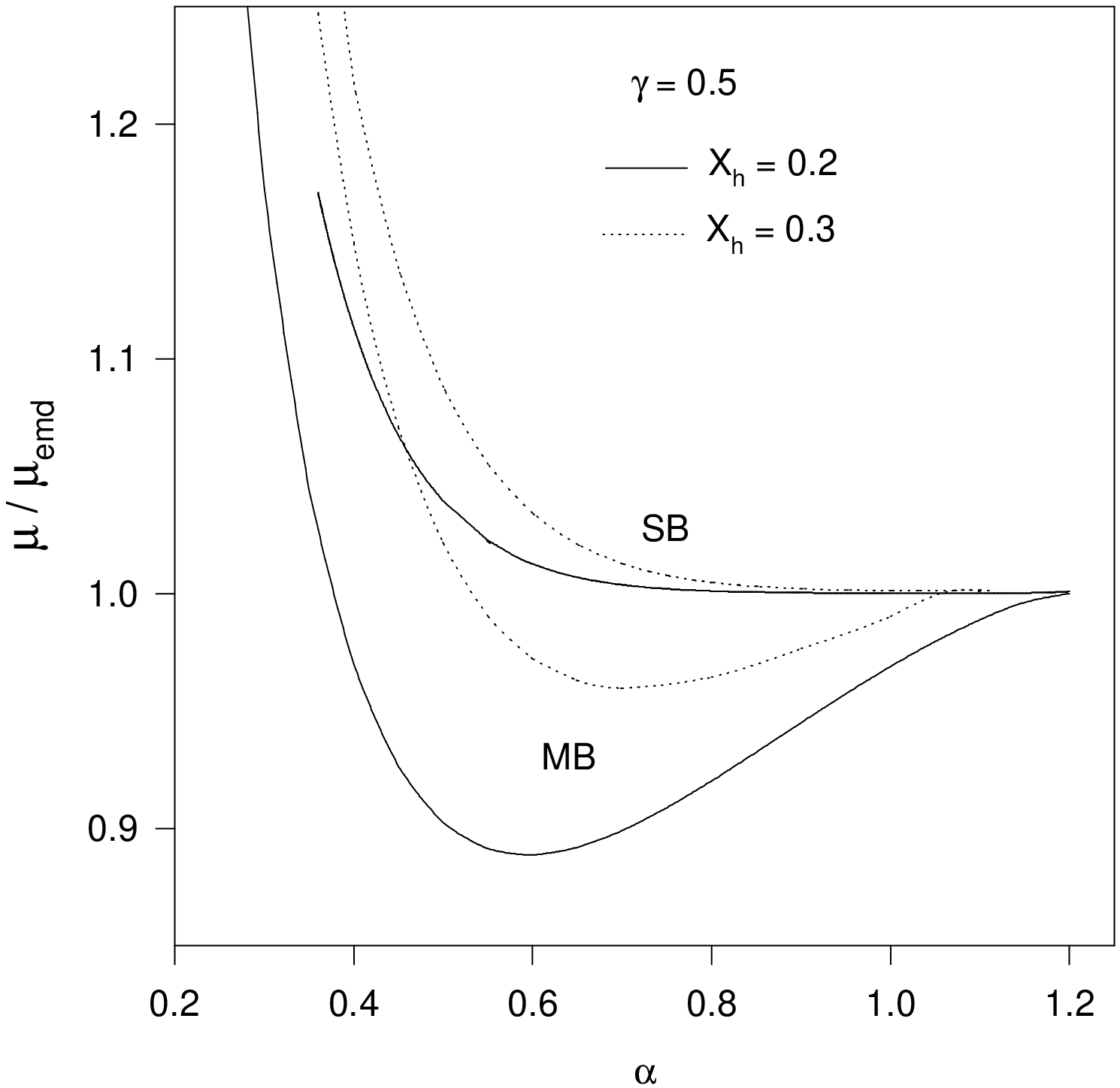}}
\caption{ The mass ratio $\mu(\infty)/\mu_{\rm EMD}(\infty)$
of the non-abelian black hole
solutions and the extremal EMD solutions is shown for
$\gamma=0.5$ and for $x_h=0.2$ and $x_h=0.3$.}
\end{figure}
\end{fixy}


\vfill
\newpage

\begin{thebibliography}{99}

\bibitem{ewein}
 K. Lee, V.P. Nair and E.J. Weinberg,
 Phys. Rev. {\bf D45} (1992) 2751.

\bibitem{bfm}
 P. Breitenlohner, P. Forgacs and D. Maison,
 Nucl. Phys. {\bf B383} (1992) 357;\\
 P. Breitenlohner, P. Forgacs and D. Maison,
 Nucl. Phys. {\bf B442} (1995) 126.

\bibitem{lw}
 A. Lue and E.J. Weinberg,
 Phys. Rev. {\bf D60} (1999) 084025;\\
 Y. Brihaye, B. Hartmann, and J. Kunz,
 Phys. Rev. {\bf D62} (2000) 044008.

\bibitem{thooft}
 G. `t Hooft,
 Nucl.~Phys.~ {\bf B79} (1974) 276;\\
 A.~M. Polyakov,
 JETP Lett. {\bf 20} (1974) 194.

\bibitem{frie}
 J.~A. Frieman and C.~T. Hill,
 SLAC-Report No. SLAC-PUB-4283, 1987 (unpublished).

\bibitem{foot0}
 For vanishing and small Higgs mass $\alpha_{\rm cr} \le \alpha_{\rm max}$
 \cite{bfm}, while for large Higgs mass the Lue-Weinberg phenomenon
 occurs \cite{lw}.

\bibitem{aichel}
 P.~C. Aichelburg and P. Bizon,
 Phys. Rev. {\bf D48} (1993) 607.

\bibitem{ash}
 A. Ashtekar, A. Corichi, and D. Sudarsky
 Class. Quant. Grav. {\bf 18} (2001) 919.

\bibitem{sud}
 D. Nunez, H. Quevedo, and D. Sudarsky,
 Phys. Rev. Lett. {\bf 76} (1996) 571.

\bibitem{ymhd}
 P. Forgacs and J. Gyueruesi,
 Phys. Lett. {\bf B366} (1996) 205.

\bibitem{hl}
 J.~A. Harvey and J. Liu,
 Phys. Lett. {\bf B268} (1991) 40;\\
 J.~P. Gauntlett, J.~A. Harvey and J. Liu,
 Nucl. Phys. {\bf B409} (1993) 363.

\bibitem{emd}
 G.~W. Gibbons and K. Maeda,
 Nucl. Phys. {\bf B298} (1988) 741;\\
 D. Garfinkle, G.~T. Horowitz and A. Strominger,
 Phys. Rev. {\bf D43} (1991) 3140.

\bibitem{kks-long}
 B. Kleihaus, J. Kunz and A. Sood,
 Phys. Rev. {\bf D54} (1996) 5070.

\bibitem{jul}
 B. Julia and A. Zee,
 Phys. Rev. {\bf D11} (1975) 2227.

\bibitem{bhk}
 Y. Brihaye, B. Hartmann, J. Kunz,
 Phys. Lett. {\bf B441} (1998) 77;\\
 Y. Brihaye, B. Hartmann, J. Kunz, and Nadege Tell,
 Phys. Rev. {\bf D60} (1999) 104016.

\bibitem{foot1}
 The equations of motion are invariant under a shift
 $\psi \rightarrow \psi + \psi_0$,
 together with a rescaling $x\rightarrow x e^{\gamma \psi_0}$.
 Therefore solutions regular at infinity can always be chosen to satisfy
 $\psi(\infty)=0$.

\bibitem{foot2}
 For larger values of $\beta$, the non-abelian monopole solutions
 do not bifurcate with a branch of abelian solutions at $\alpha_{\rm cr}$.
 Instead the critical solution is essentially non-abelian \cite{lw}.

\bibitem{foot3}
 Similarly, Einstein-Yang-Mills-dilaton solutions tend to an
 Einstein-Maxwell-dilaton solution (in the limit of large node number)
 \cite{kks-long},
 while Einstein-Yang-Mills solutions tend to an
 Einstein-Maxwell (RN) solution.

\bibitem{foot4}
 For instance, for $\alpha=1$ we find $\gamma_{\rm max}(\alpha=1)=0.9872$,
 which is very close to $\alpha_{\rm max}(\gamma=1)=0.9882$.

\bibitem{lm}
 G. Lavrelashvili and D. Maison,
 Nucl. Phys. {\bf B410} (1993) 407.

\bibitem{foot5}
 For $\gamma=0.5$, the determination of $\alpha_{\rm max}$
 becomes numerically unreliable for $\beta > 6$,
 for $\gamma=1$, the analysis becomes problematic already for $\beta=3$.

\bibitem{foot6}
 With increasing $\beta$, the dilaton function $\psi(x)$
 develops a minimum close to the origin,
 with strongly increasing curvature.

\bibitem{foot7}
 We expect, that for $\gamma \rightarrow 1.4088$,
 the maximal value of $\gamma$
 for which regular solutions of YMHD theory exist,
 the domain of existence shrinks to zero size.

\bibitem{hkk}
 B. Hartmann, B. Kleihaus, and J. Kunz,
 Phys. Rev. Lett. {\bf 86} (2001) 1422;\\
 B. Hartmann, B. Kleihaus, and J. Kunz,
 Axially symmetric monopoles and black holes in Einstein-Yang-Mills-Higgs theory,
 hep-th/0108129.

\bibitem{ewein2}
 S.~A. Ridgway and E.~J. Weinberg,
 Phys. Rev. {\bf D52} (1995) 3440.

\bibitem{COLSYS}
 U. Ascher, J. Christiansen, R.~D. Russell,
 A collocation solver for mixed order systems of boundary value problems,
 Mathematics of Computation {\bf 33} (1979) 659;\\
 U. Ascher, J. Christiansen, R.~D. Russell,
 Collocation software for boundary-value ODEs,
 ACM Transactions {\bf 7} (1981) 209.

\end{thebibliography}
\end{document}